\documentclass[aps,prd,english,superscriptaddress,11pt,notitlepage]{revtex4}
	\usepackage[colorlinks=true, a4paper=true, pdfstartview=FitV,
linkcolor=blue, citecolor=blue, urlcolor=blue]{hyperref}

\usepackage{amsmath}
\usepackage{amssymb}
\usepackage{graphicx}
\usepackage{hhline}
\usepackage{textcomp}
\makeatletter
\usepackage{babel}
\newcommand{\bea}{\begin{eqnarray}}
\newcommand{\eea}{\end{eqnarray}}

\newcommand{\be}{\begin{equation}}
\newcommand{\ee}{\end{equation}}
\usepackage[active]{srcltx}
\begin{document}


\title{Domain walls that do not get stuck on impurities
       \vspace*{12mm}}

\author{C. Adam}
\affiliation{Departamento de F\'isica de Part\'iculas, Universidad de Santiago de Compostela and Instituto Galego de F\'isica de Altas Enerxias (IGFAE) E-15782 Santiago de Compostela, Spain}

\author{K. Oles}
\affiliation{Institute of Physics,  Jagiellonian University, Lojasiewicza 11, Krak\'{o}w, Poland}
\author{T. Romanczukiewicz}
\affiliation{Institute of Physics,  Jagiellonian University, Lojasiewicza 11, Krak\'{o}w, Poland}
\author{A. Wereszczynski}
\affiliation{Institute of Physics,  Jagiellonian University, Lojasiewicza 11, Krak\'{o}w, Poland}

\begin{abstract}
We present a field theoretical model which allows for domain walls that do not get stuck on impurities. 
For this purpose, we consider a generalized chiral magnet domain wall action with an impurity which couples both to a potential term as well as to the Dzyaloshinskii-Moriya interaction energy. We show that in the so-called critical coupling limit the interaction energy between two types of domain walls and the impurity vanishes. Therefore, at low velocity, such domain walls pass freely through the impurity with almost no loss of energy. 
\end{abstract}

\maketitle

       \vspace*{0.2cm}

\section{Introduction}
Chiral magnets are well-known to support topologically stable spin structures which have stirred up a lot of attention recently because of their possible use as information storage devices \cite{storage1}-\cite{storage2}. More concretely, for an effectively planar system the relevant structure (topological soliton) is known as a magnetic Skyrmion \cite{bogdanov1}, whereas for an effectively one-dimensional system the relevant soliton is the magnetic domain wall. In particular, possible memory devices based on magnetic domain walls are known under the names of domain wall memory (DWM) or racetrack memory \cite{retrack1}-\cite{retrack2}. It was experimentally observed, however, that the motion (using the spin transfer torque mechanism \cite{slonczewski}-\cite{berger}) of these domain walls through a nanowire typically required much stronger currents than expected previously, pointing towards a rather large energy loss \cite{grollier}-\cite{koyama}. It was soon understood that this energy loss is caused by the interaction of the domain walls with impurities which are unavoidably present in the wire. Simply, domain walls get stuck at impurities. On the other hand, magnetic textures of the skyrmionic topology can be manipulated with charge current of very low densities \cite{sk-low-current}. Furthermore, while moving in the medium, skyrmions have a tendency to avoid impurities \cite{rosch} which results in a current-velocity almost independent of the impurity pinning \cite{iwasaki}. Of course, this property is crucial for the transport of Skyrmions i.e., transport of information, through wires. 

Chiral magnets are described by the magnetisation vectors $\vec m (\vec x_i)$ and their interactions, where $\vec x_i$ are the coordinates of the atomic sites. It is customary, however, to describe the magnet in terms of an effective field theory for a magnetisation field $\vec m (\vec x)$, where the (discrete) sites $\vec x_i$ are replaced by the continuous coordinates $\vec x$ (and sums by integrals). The resulting action functional (or energy functional in the static case) then provides the basic quantity for the study of magnets, the formation of topological structures, as well as their interaction with impurities.  

If a generic impurity is added to a domain wall action, this strongly influences the properties of the domain walls (kinks) - see for example \cite{kivshar1}-\cite{ekomasov3}.  The impurity explicitly breaks the translational invariance of the model and, therefore, solitons cannot be freely shifted from one location within the wire to another. Hence, there is a non-trivial interaction between the original (no-impurity) kinks and the impurity, which can be attractive or repulsive, leading to the existence or non-existence of soliton-impurity bound states. 
In other words, solitons which move through a chain of impurities get stuck, either on one of the impurities, for an attractive interaction, or between two of them, for a repulsive interaction. In any case, to manipulate the domain wall, i.e., to change its position, one has to add a significant amount of energy. Obviously, a domain wall cannot avoid the impurities, as can happen for planar Skyrmions (which may move around point-like defects), and always loses a fraction of its energy while passing through them. This is the reason why domain walls in magnetic systems typically require a significant driving current for their motion, which limits their application as information carriers. 

From a mathematical point of view, this behavior is due to the lack of the Bogomolnyi-Prasad-Sommerfield (BPS) property \cite{bogomolny}-\cite{prasad} after the inclusion of an impurity. This means that the kinks are no longer solutions of a pertinent {\it first} order, so-called Bogomolnyi equation but, instead, solve the full second order Euler-Lagrange equations. 
  As a consequence, they do not saturate the topological bound on the  energy, and a non-trivial force between kinks and the impurity emerges. 

Recently it has been shown, however, how to couple an impurity to a solitonic system in such a way that half of the kinks keep the BPS property \cite{BPS-imp-1}. Physically, this implies that the binding energy between the kink (domain wall) and the impurity is zero, and the kink may be translated freely through the wire, although it changes its shape during this translation \cite{BPS-imp-2}. This property is related to the existence of a generalised translational symmetry, which replaces the normal translational symmetry of the system without impurities.
It is the aim of the present work to apply these ideas to chiral magnets and obtain domain walls for which, at low speed, impurities are effectively transparent. 

\section{Generalized Chiral Magnet}
The phenomenological energy functional of a chiral magnet where symmetry is broken in the $z$-direction has the following expression \cite{bog1}
\be
\hspace*{-0.5cm} E_{\chi}= \int d^3x \left[ \frac{J_s}{2} \nabla \vec{m} \cdot \nabla \vec{m}  + \mathcal{U} \right] +  E_{DM}^\alpha ,
\ee
where $\vec{m} \in \mathbb{S}^2$ is the unit length three-component magnetisation vector. The first term (quadratic in derivatives) is the Dirichlet (or Heisenberg) energy, while the second term 
\be
\mathcal{U} =  K(1-m_z^2) + \mu_0 HM (1-m_z)
\ee
is a potential $\mathcal{U}$ (a non-derivative part) and describes both the anisotropy $K$ of the system and an external magnetic field $H$. Here, $\mu_0$ is the magnetic permeability of the vacuum  and $M$ the saturation magnetization. The last term is the Dzyaloshinskii-Moriya interaction energy \cite{DM}-\cite{moriya}. In its most general form, it reads \cite{Sch-DM}
\bea
E_{DM}^{\alpha}= \sqrt{2} \kappa \int d^3 x  \left[ \cos \alpha ( m_x\partial_y m_z - m_y \partial_x m_z + m_z (\partial_x m_y -\partial_y m_x) ) \right. &+&  \\ 
\left. \sin \alpha (-m_x\partial_x m_z - m_y \partial_y m_z + m_z (\partial_x m_x +\partial_y m_y) ) \right]
\eea
where $\alpha$ is a parameter. For example, for $\alpha=0$ the DM energy reads $E_{DM}=\kappa \vec{m} \cdot \nabla^{(2)} \times \vec{m}$. $J_s$ and $\kappa$ are coupling constants (the factor of $\sqrt{2}$ in front of $\kappa$ is introduced for later convenience). 

It is a well-known fact that this model admits domain wall solutions. We assume a general form of the magnetic vector field which solves the target space constraint $\vec m^2 =1$, 
\be
\vec{m}=(\sin \phi \cos \psi, \sin \phi \sin \psi, \cos \phi)
\ee
where $\phi, \psi$ are the angles on the target space two-sphere. 
We insert it into the energy and assume that the scalar fields depend only on the $x$ direction. Hence, the domain wall energy (per unit area) reads
\be
\hspace*{-0.0cm} E_{\chi}=\int d x \left[ \frac{J_s}{2} (\phi_x^2 + \sin^2 \phi \psi_x^2) + \sqrt{2} \kappa \left(\phi_x \sin(\alpha+\psi) + \frac{1}{2} \sin 2\phi \cos (\alpha+\psi)\psi_x\right) + \mathcal{U} \right] .
\ee
The minimization of the energy functional requires $\delta_\psi E=0$, which leads to the EL equation solved by a constant 
\be
\psi_0=\frac{\pi}{2} -\alpha
\ee
which describes the interpolation between Bloch domain walls ($\alpha=0$) and N\'{e}el domain walls ($\alpha=\pi/2$). For intermediate values of $\alpha$, we have a mixture between these two cases, where the magnetization vector rotates in all three dimensions. 
Hence, only the $\phi$ field has a nontrivial spatial dependence, which is given by the following effective energy 
\be
E_{\chi}=\int d x \left[ \frac{J_s}{2} \phi_x^2   + \sqrt{2} \kappa   \phi_x  +\mathcal{U}(\phi)  \right]  .\label{phi-en}
\ee
Of course, the DM interaction energy no longer affects the field equations, because it is a total derivative. 

Now we want to consider a modification of this energy functional which can host an impurity $\sigma(x)$ without completely breaking the BPS property. First of all, we assume that the DM coupling constant is promoted to a spatially dependent function $\kappa=\kappa(x)$. Hence, we get a term $\sqrt{2} \kappa(x) \phi_x$. Now this cannot be trivially reduced to a boundary term and, therefore, does have a nontrivial impact on the field equation. Such a modification of the DM term could either be an effect of the appearance of the impurity, or related to another, hopefully experimentally controlled, mechanism. Secondly, the impurity couples to a potential-like term $W(\phi)$. Observe that this is the usual way of how an impurity enters a physical system \cite{kivshar1}-\cite{kivshar2}. Thus, we arrive at (we assume $J_s=1$)
\be
E=\int_{-\infty}^\infty dx \left[  \frac{1}{2} \phi_x^2 +\sqrt{2} \kappa(x) \phi_x+\mathcal{U}+2 \sigma (x) W +\sigma^2(x) \right] 
\ee
where we also added a self-energy part for the impurity. This addition only sets the energy scale and does not enter the equation of motion. Next, we consider a critical coupling limit of this model in which the following relations hold
\be
W^2=\mathcal{U} \;\;\;\; \mbox{and} \;\;\;\; \kappa (x)= \sigma(x) + \kappa_0.
\ee
Here, $\kappa_0$ is the value the DM coupling constant would have in the absence of the impurity.
Hence, we arrive at
\be
E_{BPS}=\int_{-\infty}^\infty dx \left[ \frac{1}{2}\phi^2_x +\sqrt{2} \left( \sigma (x) + \kappa_0 \right) \phi_x +W^2 + 2\sigma (x) W  +\sigma^2 (x) \right] . \label{BPS}
\ee
We keep $W$ as the primary ingredient. It is usually referred to as a prepotential. Here BPS means that, in contrast to the generic impurity chiral magnet model with the domain wall geometry, this particular model possesses a topological energy bound which is saturated by solutions of the pertinent first order Bogomolnyi equation \cite{BPS-imp-1}.  Therefore, there will exist soliton-impurity solutions with {\it zero binding energy}, which results in zero static forces between these solitons and the impurity. This happens for any form of the impurity and any (at least double vacuum) potential $W^2$, as we will show in the next section. 

Of course, the main question is whether such a generalization of the chiral magnet can be realized experimentally. 
\section{The half-BPS kink impurity model}
Let us prove that the energy (\ref{BPS}) possesses a saturated topological energy bound. For that we write it as a full square
\be
 E =\int_{-\infty}^{\infty} dx \left( \frac{1}{\sqrt{2}} \phi_x + (\sigma + W)\right)^2  -\sqrt{2} \int_{-\infty}^{\infty} dx \phi_x (W - \kappa_0) \geq   - \sqrt{2} \int_{\phi (-\infty)}^{\phi (+\infty)} d\phi (W - \kappa_0) 
 \ee
 where the last expression is a topological one, that is, it depends only on the asymptotic (vacuum) values of the field an not on its particular local shape. The bound is saturated if and only if the following Bogomolnyi equation is satisfied
 \be
  \frac{1}{\sqrt{2}} \phi_x + \sigma + W=0.
 \ee
 It is straightforward to show that this equation implies the static  EL equation. Hence, its solutions obey the full equations of motion. Note that, in contrast to the no-impurity case, there is only {\it one} Bogomolnyi equation. Indeed, for $\sigma=0$ the two first-order equations $(1/\sqrt{2}) \phi_x \pm W =0$ may be derived. As a consequence, here only half of the solitons are of the BPS type. 
 
 It can also be shown that the topologically trivial solution, (the {\it lump}), obeys the Bogomolnyi equation. Its energy is, therefore, 0 and it is a counterpart of the vacuum in the no-impurity model. 
 Observe that the BPS property occurs for any (admissible) $W$ and $\sigma$. Hence, there is no need for any fine tuning at this stage. 
 
 Furthermore, it can be rigorously proved that in the BPS sector there exists a generalized translation symmetry which means that there are infinitely many energetically equivalent solutions describing a topological soliton at an arbitrary distance from the impurity \cite{BPS-imp-2}. Of course, the form (shape) of the solution depends on the distance between the soliton and the impurity. This results in the existence of a zero mode \cite{BPS-imp-2}. When the soliton moves, it is no longer a BPS state. However, at low speed the dynamics of the domain wall may be approximated by a sequence of BPS states, i.e., a geodesic motion on the moduli space, which implies an almost elastic interaction with the impurity and, therefore, arbitrarily small energy loss. In conclusion, the BPS soliton will freely pass through the impurity.
 
The generalized translation acts also on the lump. However, it transforms the lump solution into itself. Obviously, the lump is confined to the impurity and reflects a modification of the original vacuum due to the insertion of the BPS impurity. 
\section{The sine-Gordon model with the BPS impurity}
Let us turn back to the chiral magnet model and take the potential relevant for this case. For reasons of simplicity we neglect the asymmetry term, assuming $K=0$. Hence,
\be \label{K=0}
\mathcal{U}=\beta (1-\cos \phi)=2\sin^2 \frac{\phi}{2}
\ee
which is the famous sine-Gordon potential. Here, $\beta=\mu_0 HM=1$. For the sine-Gordon potential the choice of the prepotential is not unique. We choose it in a sign changing form 
\be
W=\sqrt{2} \sin \frac{\phi}{2} .
\ee
We want to stress that all main qualitative results hold also for the $K\neq 0$ case.
Finally, we also shall set $\kappa_0 =0$ in the rest of the paper. The Bogomolnyi and EL equations do not depend on $\kappa_0$, therefore, all solutions are the same. All that changes are the energies of kinks (antikinks) which receive a constant positive (negative) contribution linear in $\kappa_0$.
\subsection{Structure of the static sector}
First of all, this choice for the prepotential significantly affects the target space symmetry of the original model. Indeed, without impurity the sine-Gordon model is invariant under the translation $\phi \rightarrow \phi +2\pi$. Now, the impurity model only enjoys half of this symmetry, $\phi \rightarrow \phi + 4 \pi$. This means that the fundamental domain of solutions (value of the field) is $[-2\pi, 2\pi)$ and not $[0,2\pi)$, as for the sine-Gordon model.  Therefore, instead of {\it two} different domain walls of the original sine-Gordon model, we have {\it four} solitons. Similarly, the half-BPS impurity model has {\it two} topologically trivial (nonequivalent) solutions, i.e., lumps $\Sigma_0, \Sigma_{2\pi}$, which are the field responses (deformations) of the $\phi=0$ and $\phi=2\pi$ vacua of the original sine-Gordon model to the impurity insertion. 

\begin{figure}
\includegraphics[height=4.5cm]{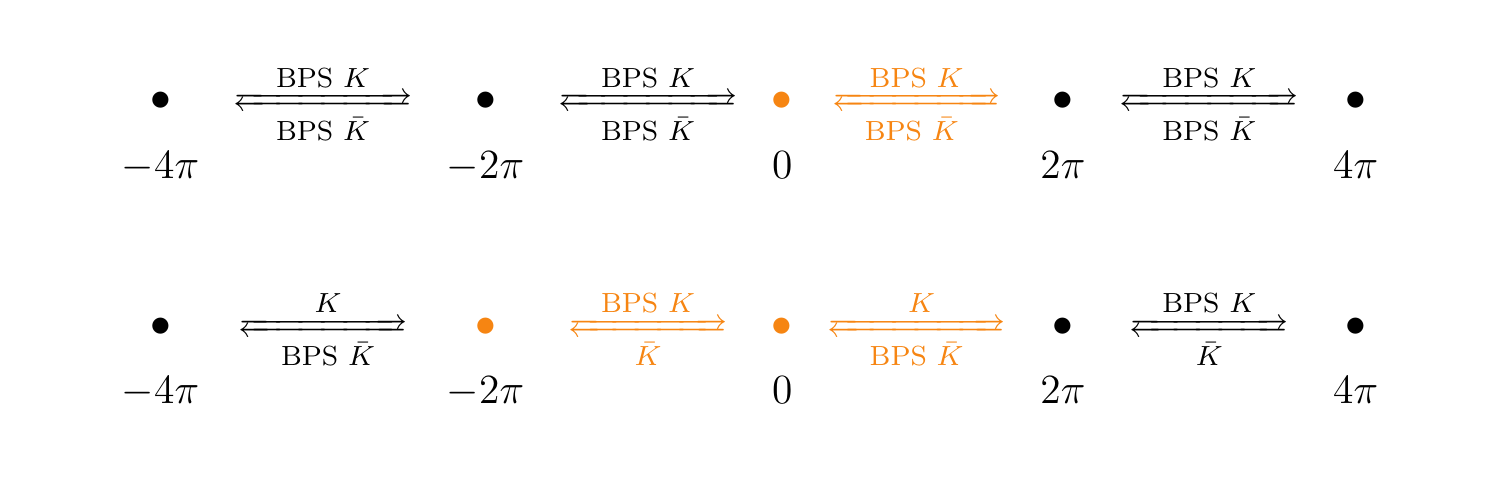}
\caption{Structure of the static soliton solutions for the original sine-Gordon model (upper line) and the fermionic impurity case (lower line). The orange dots and arrows denote the vacua and the solitons in the fundamental domain $[0, 2\pi)$ and $[-2\pi, 2\pi)$ respectively. }
\label{fermionic}
\end{figure}

Secondly, the BPS solutions do not have to have a fixed topological sign, that is, a kink $(Q=+1)$ as well as an antikink $(Q=-1)$ can be of the BPS type. This is possible because of the twice bigger fundamental domain, which means that we have two {\it non-equivalent} kinks (and antikinks). Let us then consider the Bogomolnyi equation 
  \be
  \frac{1}{\sqrt{2}} \phi_x = - \sigma - W =-\sigma - \sqrt{2} \sin \frac{\phi}{2}.
 \ee
As the impurity is assumed to be exponentially localized (or at least very well localized) the {\it asymptotical} sign of the field derivative is governed (except for the critical cases considered below) by the sign of the prepotential. 
In the fundamental domain, $W>0$ for $\phi \in [0,2\pi]$ and $W<0$ for $\phi \in [-2\pi, 0]$. This implies that we have a BPS antikink interpolating between $2\pi \rightarrow 0 $ and a BPS kink interpolating between  $-2\pi \rightarrow 0$. On the other hand, a kink interpolating between $0 \rightarrow 2\pi$ and an antikink between $0 \rightarrow -2\pi$ cannot be of the BPS type. The structure of the static soliton solutions is presented in Fig. \ref{fermionic}. 

Note that in the original sine-Gordon model the shift $\phi \rightarrow \phi +2\pi$, if applied to the unit vector $\vec{m}$, is a rotation on the target $\mathbb{S}^1 \subset \mathbb{S}^2$. Hence, it rotates the vector back into its original position. Here, after adding the impurity with the prepotential of the first type, a physically equivalent configuration is obtained after a $4\pi$ rotation. This is a characteristic feature for fermions. Therefore we call this case a {\it fermionic impurity}.

The energy of the BPS solutions (the kink and antikink) for the fermionic impurity is the same $E_{K_{BPS} } = E_{\bar{K}_{BPS} } = 8$. Obviously, this does not depend on a particular choice of the functional form of the impurity (provided the solution exists). 

One should be aware that the BPS solitons do not exist for an arbitrary impurity. If the impurity is too strong, there is no solution interpolating between the vacua. To demonstrate this, we specify the impurity as
\be
\sigma = \frac{\alpha}{\cosh x}  \label{imp}
\ee
(it should be emphasized, however, that our results are qualitatively independent of the particular choice of the impurity).
\begin{figure}
 \includegraphics[height=4.8cm]{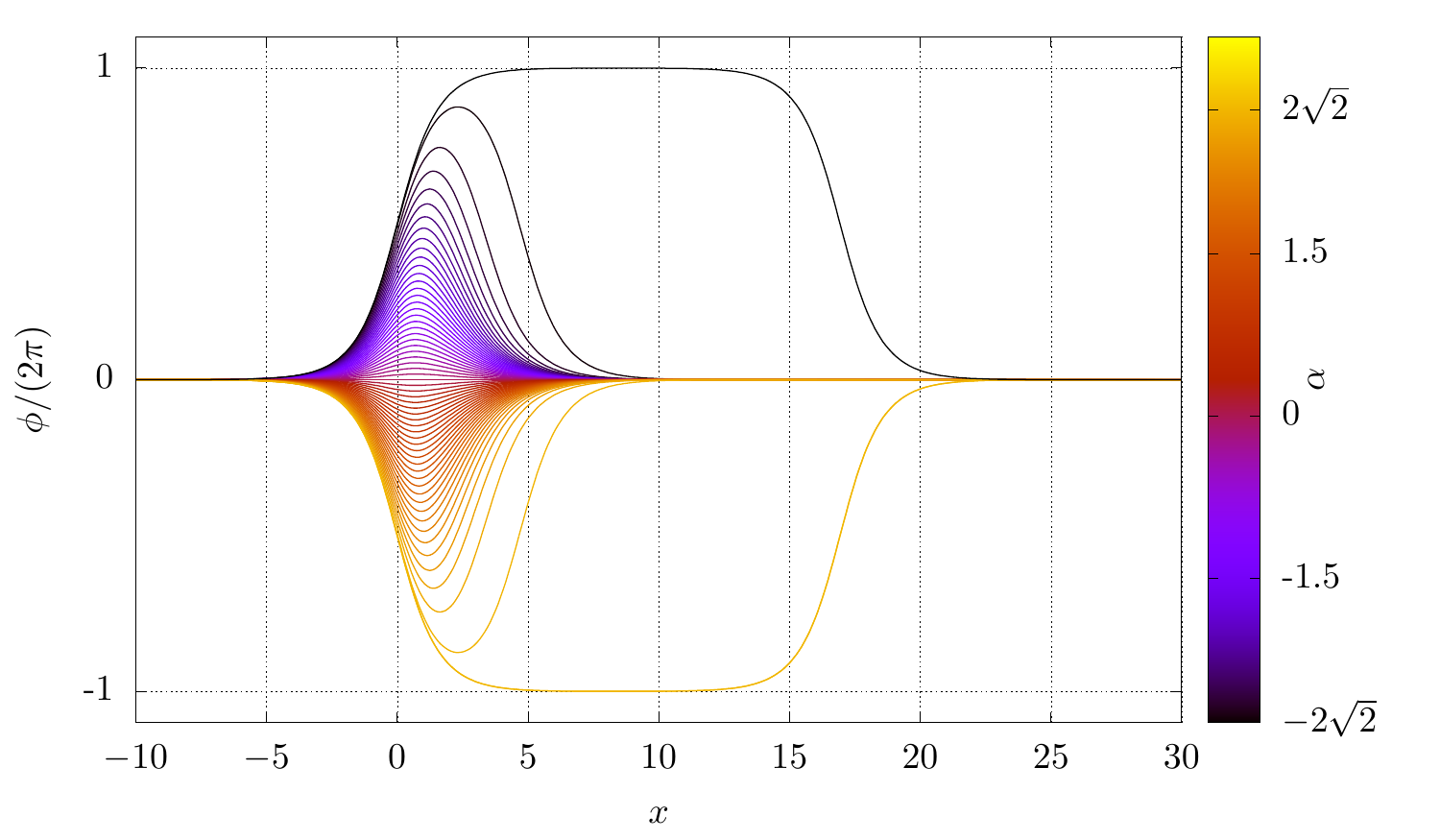}
\includegraphics[height=4.6cm]{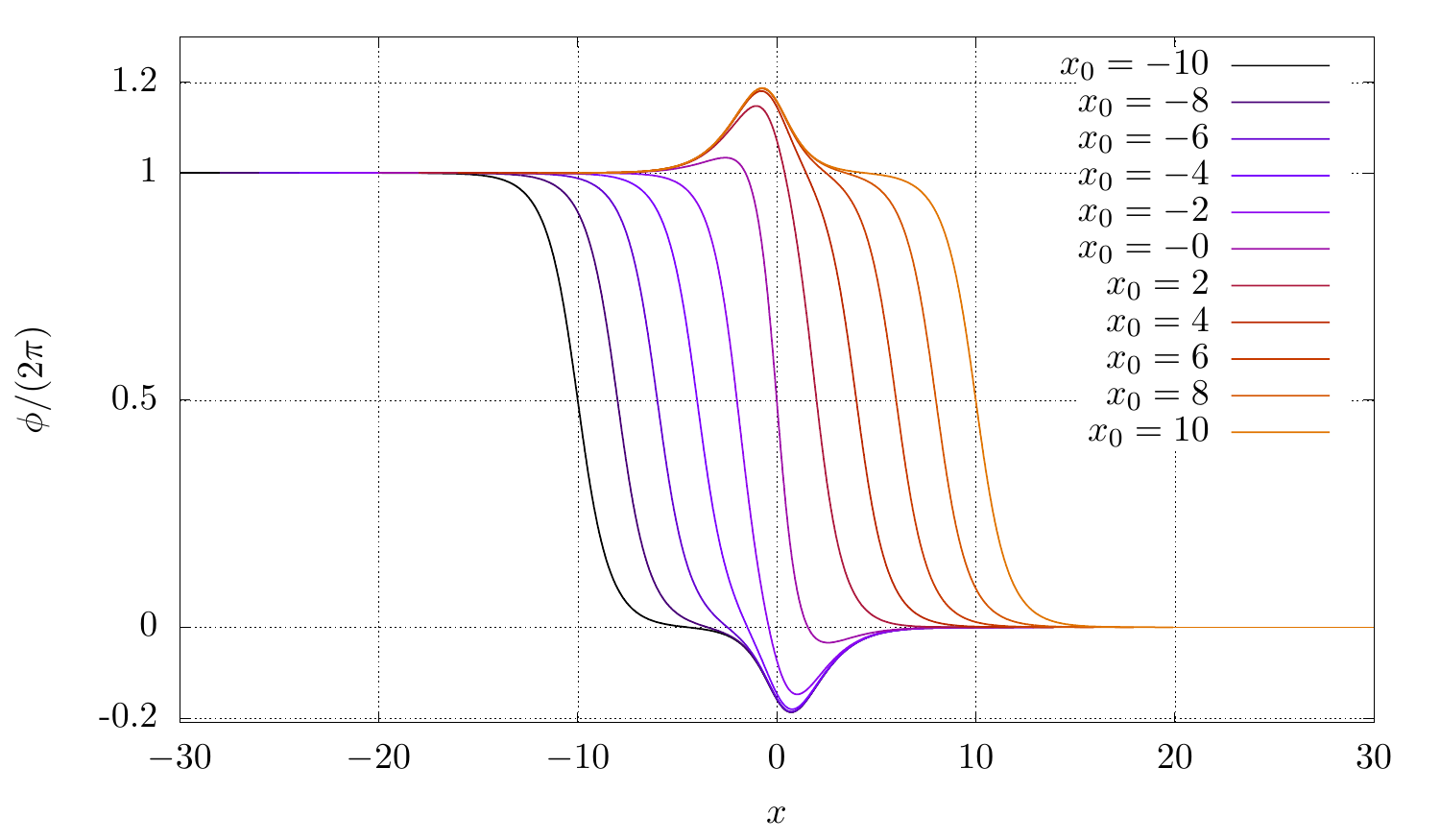}
\caption{Left: lump solutions for different values of the impurity strength $\alpha$. Right: BPS kink solutions at different distances from the impurity, for the value $\alpha =0.5$.}
\label{lump plot}
\end{figure}
(\ref{imp})  is an exponentially localized impurity centered at $x=0$, whose strength is measured by the parameter $\alpha$. In Fig. \ref{lump plot}, left panel, we plot the lumps for different values of the parameter $\alpha$. As the strength, $|\alpha|$, of the impurity grows, the field can be forced to approach one of the vacua. Because of the periodicity of the vacuum it can happen for both, positive and negative $\alpha$. Indeed, it is clearly visible that if $\alpha \rightarrow -2\sqrt{2}$ or $2\sqrt{2}$ the lump looks like a bound state of a pair of infinitely separated sine-Gordon kink and antikink. Strictly speaking, for $\alpha \rightarrow -2\sqrt{2}$ the sine-Gordon kink $\phi=4\arctan e^{x}$ is confined to the impurity while the sine-Gordon antikink $\phi= 4\arctan e^{-(x+x_0)}$ is expelled to infinity. In the limit $\alpha \rightarrow 2\sqrt{2}$, it is the sine-Gordon antikink $\phi=4\arctan e^{-x} -2\pi$ which stays at the impurity. 
This limit is also present in the Bogomolnyi equation, which for $\alpha \rightarrow \alpha_{crit}^{\pm} = \pm 2\sqrt{2}$ has analytical solutions
\be
\phi_K=4\arctan e^{x} \;\;\; \mbox{or} \;\;\; \phi_{\bar{K}}=4\arctan e^{-x} - 2\pi ,
\ee
that is, the sine-Gordon kink (antikink) localized at the impurity, while the charge conjugate counterpart (forming together the topologically trivial solution) is located at infinity. To conclude, an acceptable impurity requires $\alpha \in [-2\sqrt{2}, 2\sqrt{2}]$. 
In Fig. \ref{lump plot}, right panel, we plot the BPS antikink at different distances from the impurity (\ref{imp}) for $\alpha=0.5$. Owing to the generalized translation symmetry, all these states have the same energy and belong to the moduli space of the BPS solutions. 

\vspace*{0.2cm}

\begin{figure}
\hspace*{-1.0cm}
\includegraphics[height=6.0cm]{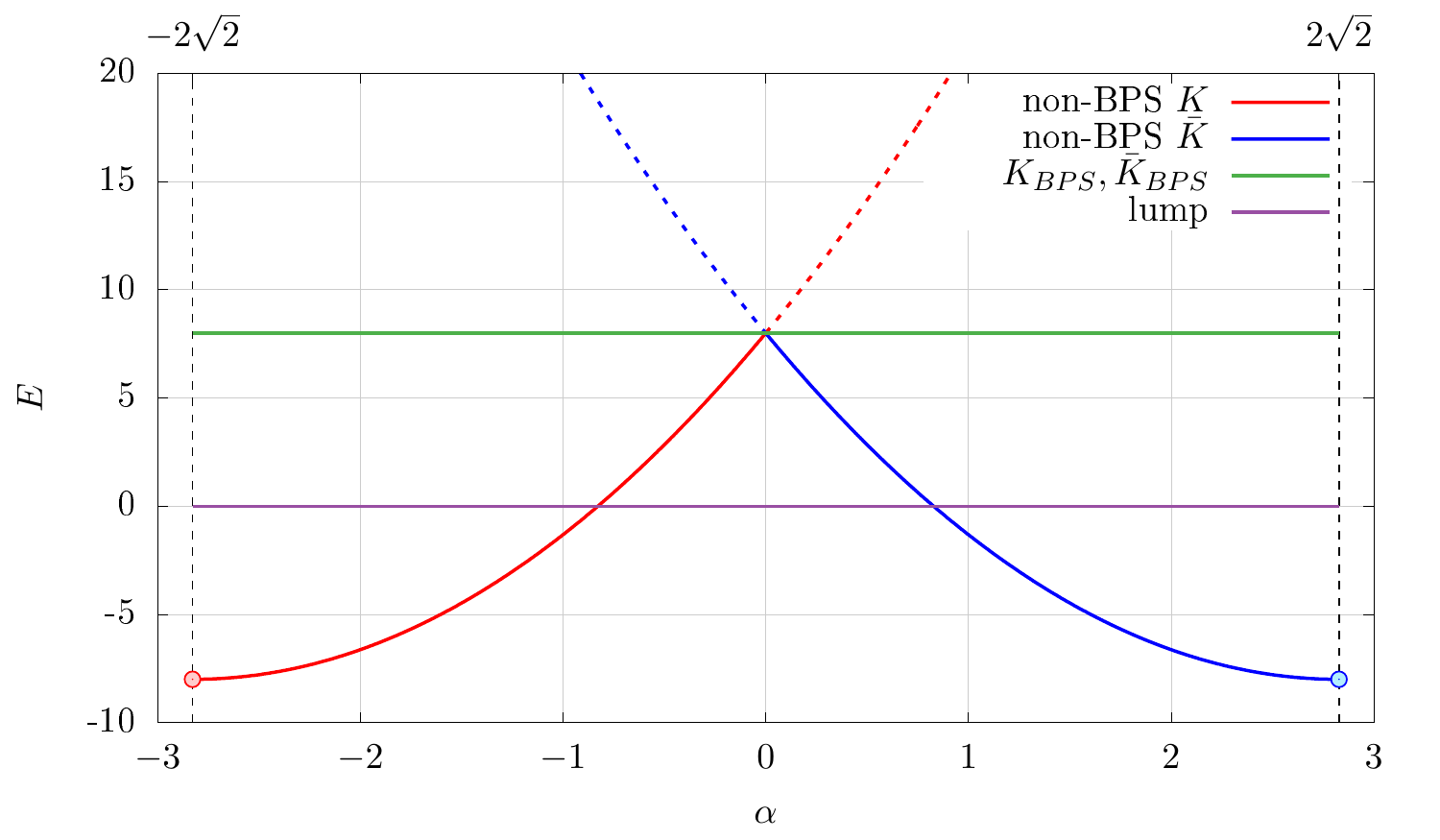}
\caption{Energy of the static solutions as a function of the impurity strength $\alpha$.}
\label{energy plot}
\end{figure}

It results that for our impurity (all acceptable $\alpha$) the non-BPS solutions can be found in an exact form given by
\be
\phi_K=4\arctan e^{x} \;\;\; \mbox{or} \;\;\; \phi_{\bar{K}}=4\arctan e^{-x} - 2\pi .
\ee
Note that they coincide with the solutions of the Bogomolnyi equation at the critical coupling strength of the impurity. However, they obey the EL equation for any $\alpha$. The energy computed for these non-BPS solutions reads
\be
 E_K=8 +8\sqrt{2} \alpha + 2\alpha^2 \;\;\;   E_{\bar{K}}=8 - 8\sqrt{2} \alpha + 2\alpha^2 .
 \ee
These expressions allow us to analyze the stability of the non-BPS solutions towards decay into the 
separated kink (antikink) and the lump. Indeed, the energy of the infinitely separated kink (antikink) is just the energy of the sine-Gordon kink (antikink), $E_{sG}=8$, while the energy of the lump is $E_{lump}=0$. Let us consider the non-BPS kink. The stability condition is 
\be
E_K - (8+0) = 2 \alpha (4\sqrt{2} + \alpha) \leq 0
\ee
This expression is always positive for positive $\alpha$ and negative for negative $\alpha$ within the acceptable values. Hence, the non-BPS kink forms a stable static bound state with the impurity if $\alpha <0$. Furthermore, it qualitatively explains the fact that the non-BPS kink is repelled from the impurity for $\alpha >0$ and attracted for $\alpha <0$, as will be seen in the next subsection. Note that the energy of the stable non-BPS kink at the critical value $\alpha=-2\sqrt{2}$ is $-8$. Hence, in this limit, the pair of kink and antikink has its energy equal to the energy of the lump. 
\\
In the case of the non-BPS antikink, the behavior is identical up to the interchange $\alpha \rightarrow -\alpha$. Hence, the non-BPS antikink is attracted by the impurity for $\alpha >0$ and repelled for $\alpha <0$ (within the acceptable values). Similarly, the non-BPS antikink forms a stable static bound state with the impurity if $\alpha >0$. Therefore, for a given $\alpha$, only one non-BPS solution forms a stable bound state with  the impurity. Moreover, one of the non-BPS solitons is always repelled and one is always attracted by the impurity. For a summary of these results, see Fig. \ref{energy plot}.

We remark that there also exists a BPS impurity which does not change the target space shift symmetry of the original sine-Gordon model, $\phi \rightarrow \phi +2\pi$. It is defined by taking the absolute value of the prepotential, $W=\pm \sqrt{2} \left| \sin \phi /2 \right|$. Then, the fundamental domain is $(0, 2\pi]$ and the antikink or the kink are BPS solutions, respectively. The topological charge conjugate soliton is a non-BPS state, nontrivially interacting with the impurity. We call such a impurity the {\it bosonic impurity}.
\subsection{Interaction with the impurity}
For a study of time dependent processes, we require a completion of the action functional for time dependent fields. For our purposes, the exact form of this time completion is not so important, because we are mainly interested in describing the adiabatic motion of BPS solitons, i.e., their motion through a sequence of static BPS kinks (the moduli space approximation).
For simplicity, we shall assume an action quadratic in time derivatives, i.e.,
\be
S_{BPS} = \int dtdx \frac{1}{2} \dot \phi^2 - \int dt E_{BPS}.
\ee
Formally, this is a completion to a Lorenz invariant kinetic term which, thus, provides the usual second order dynamics. The front velocity (dispersion in the high-frequency limit), however, obviously is {\em not} equal to the speed of light. 
\begin{figure}
\hspace*{-1.0cm}
\includegraphics[height=10.0cm]{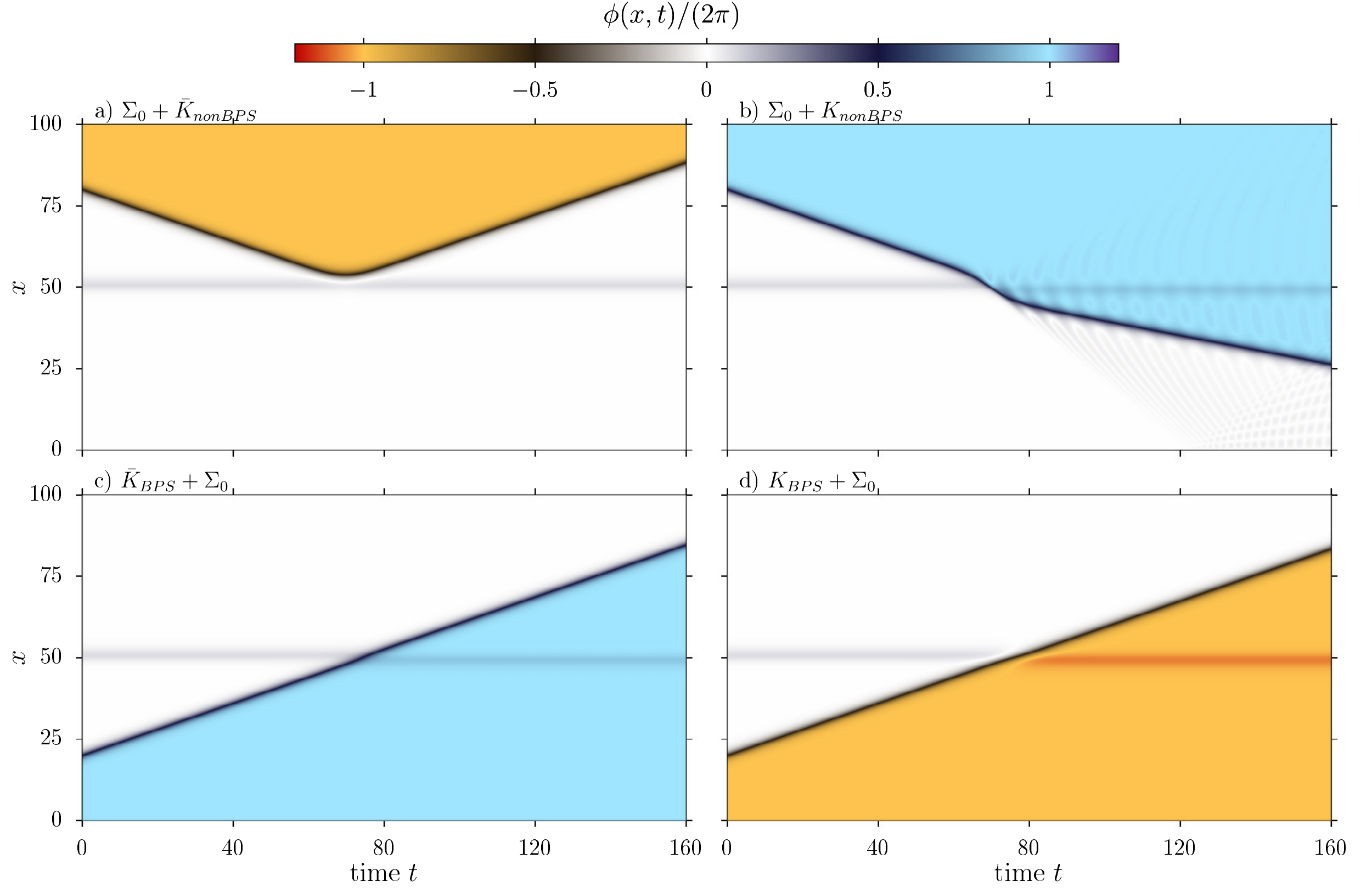}
\caption{Scattering of the solitons on the impurity with $\alpha=-0.4$.}
\label{scattering}
\end{figure}
\begin{figure}
\hspace*{-1.0cm}
\includegraphics[height=7.5cm]{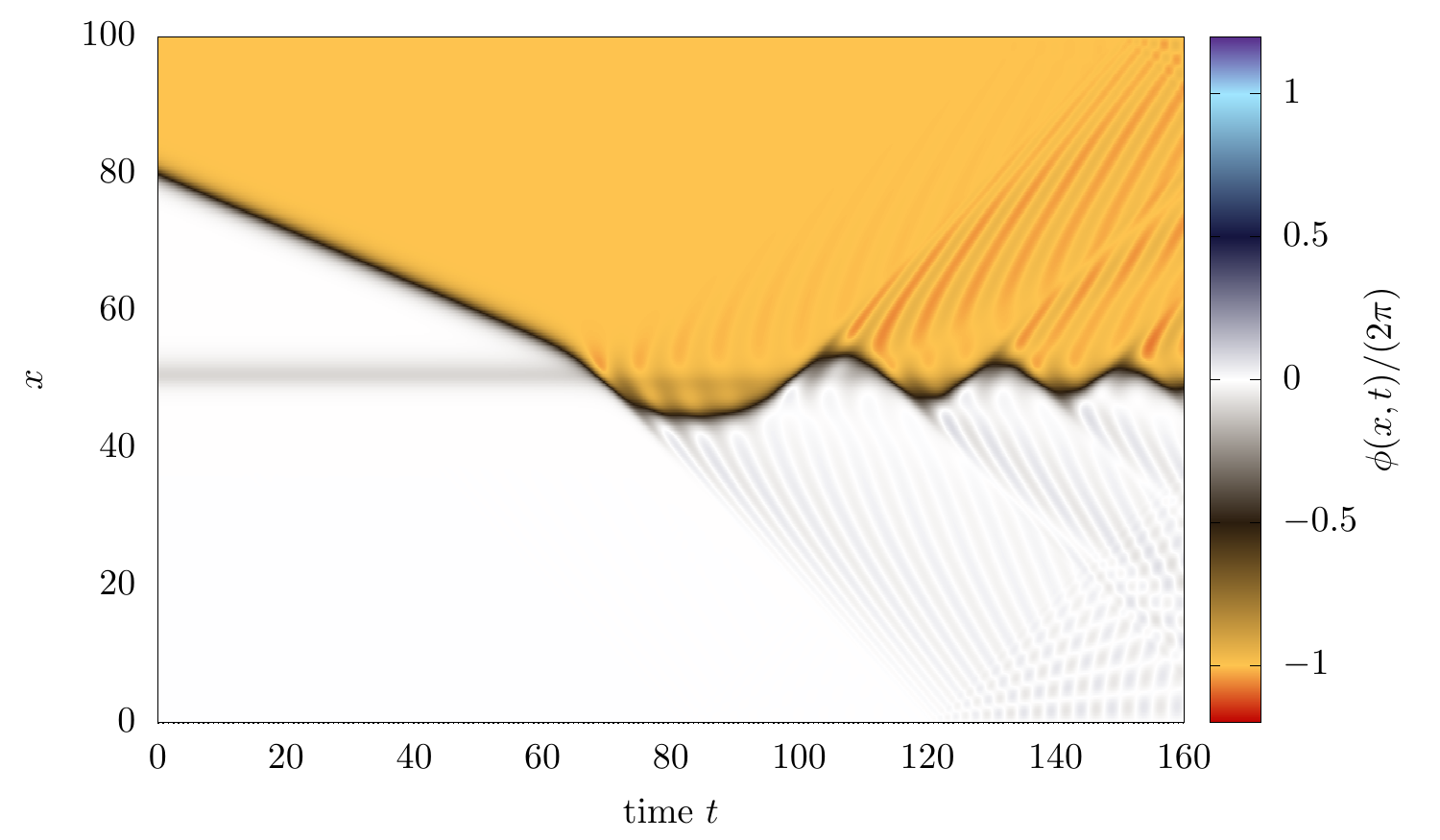}
\caption{Trapping of the antikink on the impurity. Here $\alpha=0.5$.}
\label{trap}
\end{figure}
\begin{figure}
\hspace*{-1.0cm}
\includegraphics[height=5.0cm]{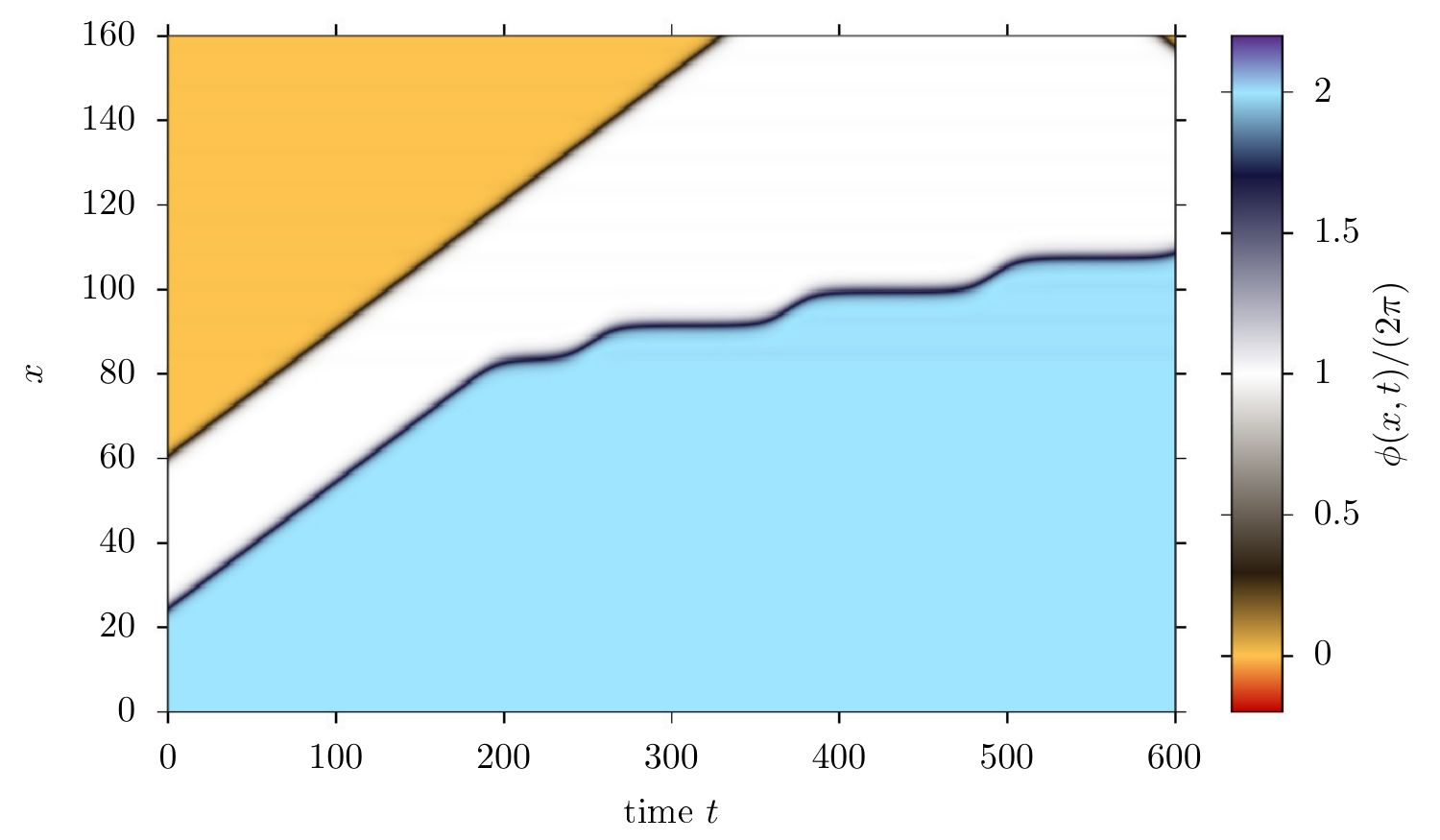}
\includegraphics[height=5.0cm]{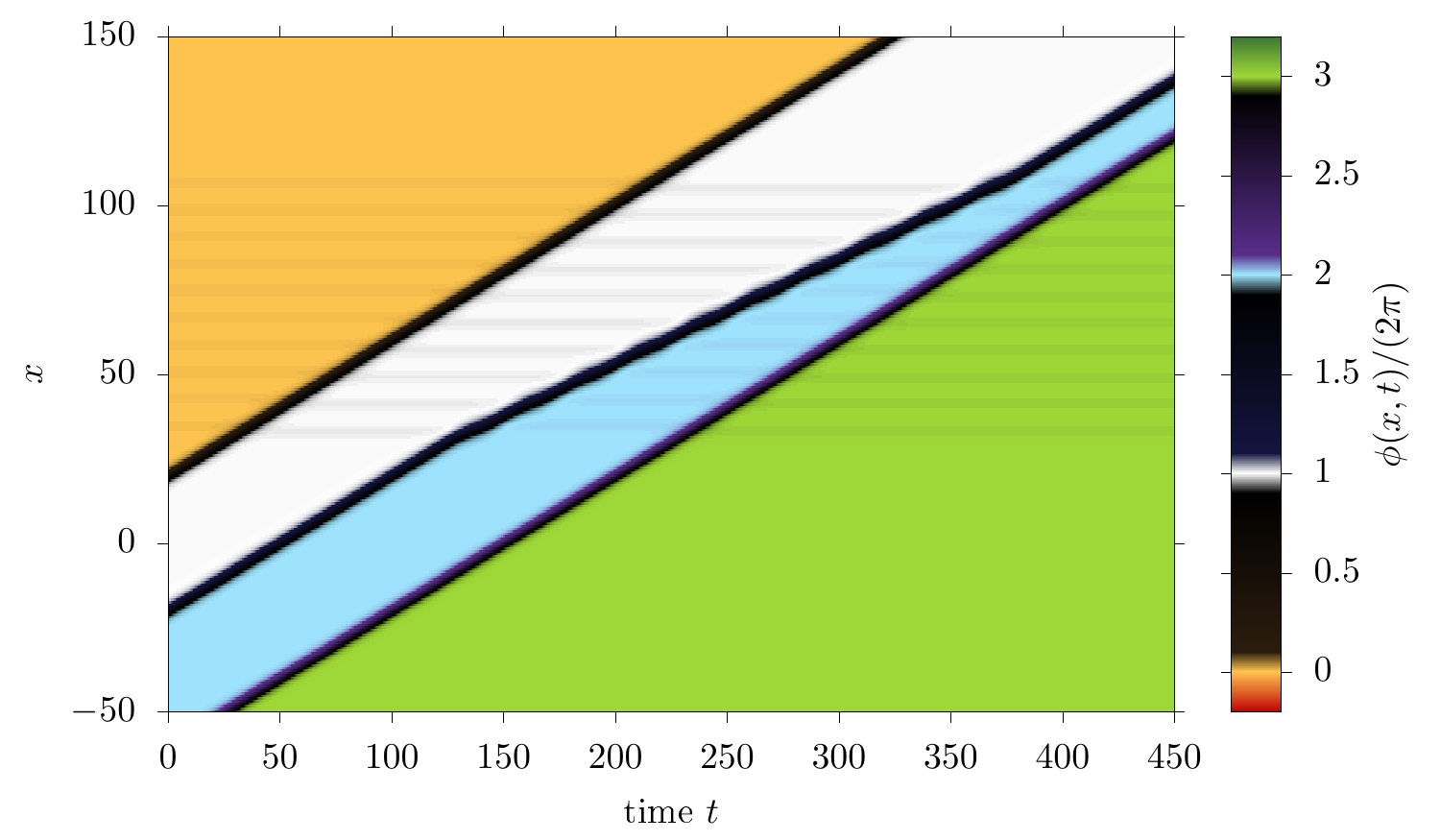}
\caption{Scattering of the antikinks on a chain of impurities.}
\label{chain}
\end{figure}

Solitons moving at a nonzero velocity are no longer BPS solutions. 
However, if the originally static BPS kink and antikink are given a sufficiently small initial velocity, they can pass through the impurity with almost no energy loss. The explanation is the existence of the moduli space, i.e., the space of all energetically equivalent static solutions of the Bogomolnyi equation with a fixed topological sector, which represent a BPS kink (or antikink) at an arbitrary distance from the impurity. Then, starting at infinity,  the motion of the slowly moving BPS kink occurs via a sequence of  subsequent BPS states.  This is very well visible in Fig. \ref{scattering}, bottom right and bottom left panels, where the BPS kink (antikink) with $v=0.4$ passes through the impurity with $\alpha=-0.4$. The initial speed of the soliton is not affected during the collision, and the energy loss is practically 0. As a consequence, one can manipulate these domain walls using a very small current. Note that the velocity used in this plot is not particularly small, which shows the robust character of the mechanism.

This should be contrasted with the scattering properties of the non-BPS solutions where significant dissipation of the initial energy of the kink is observed. In Fig. \ref{scattering}, top panels, we collide the non-BPS antikink (or kink) with $v=-0.4$ with the impurity. In the antikink collision, the soliton is strongly repelled and is almost elastically scatted back to infinity. The kink, on the other hand, is attracted by the impurity and during the collision it reduces its velocity significantly. In addition, a lot of radiation is emitted. When the impurity is strong enough or if the initial velocity is sufficiently small then the non-BPS kink can get stuck on the impurity. The trapping of the non-BPS kink is presented in Fig. \ref{trap}. Here $v=-0.4$ and $\alpha=0.5$.

If such a kink travels though a chain of impurities, it will be eventually confined to one of them - see Fig. \ref{chain}. Here, each of 10 impurities is rather weak ($\alpha=-0.033578$ and $\alpha= -0.04$ for two and three soliton case respectively) with the separation length $L=8$, while the velocity is $v=0.4$. Therefore, in order to use it as a information carrier a driving force would be needed. 
As a consequence of the structure of the static solutions, in a multi-soliton incoming state odd and even constituents behave differently. In Fig. \ref{chain} we show a two- and a three-antisoliton solution incoming from minus infinity. The first member of the multi-soliton state is the BPS antisoliton ($\phi:  2\pi \rightarrow 0$) while the second must be of non-BPS nature ($\phi: 4\pi \rightarrow 2\pi$). Hence, their interaction with the impurity differs dramatically. Note, that this effect is due to the fermionic type of the impurity. For an impurity of the bosonic type, all BPS solitons, forming an incoming multi-soliton configuration, will pass through the impurity without the energy dissipation. 

\section{Summary}
It was our main objective to demonstrate the existence of a mechanism which allows the almost dispersionless motion of magnetic domain walls through a thin wire in the presence of impurities.
In the effective field theory model describing the domain wall, this lossless motion is permitted by the existence of a first-order equation (Bogomolnyi equation) and of a generalised translational symmetry in the domain wall-impurity system. In particular, the generalised translational symmetry gives rise to the following physical picture. A BPS domain wall with a sufficiently low initial velocity starts with its standard (no-impurity) shape far away from the impurities. As it gets closer, it is distorted by the presence of the impurities, but without gaining or losing energy. Finally, sufficiently far from the impurities, it recovers its original shape. In our concrete calculations of  Section IV, we mainly considered just one exponentially localised impurity at $x=0$ (except for the scattering off a chain of impurities shown in Fig. \ref{chain}), but it is obvious from our general discussion that these results do not depend on the particular shape of the impurity. Specifically, we might as well choose an impurity function $\sigma (x)$ describing a periodic or randomly distributed pattern of individual impurities and arrive at qualitatively exactly the same conclusions. In addition to the impurity, also the potential may be chosen rather arbitrarily (of course, it should have at least two vacua, such that domain walls can exist at all). In particular, the inclusion of the asymmetry term (i.e., $K\not= 0$ in Eq. (\ref{K=0})), does not change our qualitative results. The only reason to set $K=0$ there was that the model could then be mapped onto the well-known sine-Gordon model.  

All that had to be specified was the interaction of the impurity with the field of the chiral magnet. On the one hand, the impurity couples to a potential term, which is rather standard for a coupling of impurities. On the other hand, it couples to the DM interaction in a specific way. More precisely, the DM coupling constant $\kappa_0$ is {\em modulated} by the pattern of impurities $\sigma (x)$ like $\kappa_0 \to \kappa (x) = \sigma (x) + \kappa_0$.  An important question is, of course, whether and in which way this modulation can be realised. The simplest and most elegant solution would be that the interaction of the  chiral magnet with the physical impurities by itself effectively produces this term. If this is not possible, another possibility is that the modulation of the DM coupling is introduced by engineering techniques, that is, by manipulating the material that carries the domain walls (e.g. the wire). In this latter case, it is probably not realistic that the modulation of the DM coupling exactly reproduces the impurity pattern. But it could at least do so approximately, resulting in a system which is not exactly BPS but still allows for a domain wall motion with a much reduced loss of energy. One concrete possibility, e.g., would be a periodic modulation of the DM coupling for a periodic pattern of impurities.

In order to study moving domain walls, we also had to make a choice for the time dependence of our system. For simplicity, we chose a kinetic term which is quadratic in velocities. For our main purpose, however, this choice is irrelevant. The motion of BPS domain walls a sufficiently low velocities will always be described by a motion on moduli space, that is, by an adiabatic sequence of BPS states related to each other by the generalised translational symmetry. All that may change is the range of velocities for which the moduli space approximation is valid. Of course, at higher velocities, where the dynamics can no longer be approximated by the geodesic motion on moduli space, a particular choice of time evolution (for example, defined by the Landau-Lifshitz-Gilbert equation which gives a reliable description of magnetisation dynamics \cite{lee}) may strongly affect interactions between solitons and impurities. 

Looking from a wider perspective, we find it intriguing that the DM interaction not only favours skyrmionic magnetic textures but also provides a field theoretical mechanism which (alone or together with other mechanisms as for example the perpendicular magnetic anisotropy \cite{ravelosona}-\cite{jung} and short current pulses/femtosecond laser pulses \cite{rav}) can lead to a significant reduction of the critical current density of the domain wall motion.

\section*{Acknowledgements}
The authors acknowledge financial support from the Ministry of Education, Culture, and Sports, Spain (Grant No. FPA2017-83814-P), the Xunta de Galicia (Grant No. INCITE09.296.035PR and Conselleria de Educacion), the Spanish Consolider-Ingenio 2010 Programme CPAN (CSD2007-00042), Maria de Maetzu Unit of Excellence MDM-2016-0692, and FEDER.

\end{document}